\documentclass[12pt,fleqn,a4paper]{article} % A4 size
\usepackage{epsfig,graphics,graphicx}
\usepackage{clshan-math,clshan-dm-dd}

\def \gsim {\:\raisebox{-0.7ex}{$\stackrel{\textstyle>}{\sim}$}\:}
\newcommand{\tabt}  [1] {\begin{flushleft}
                         \footnotesize
                         {\bf #1} \\
                         \vspace{0.2cm}
                         \renewcommand{\arraystretch}{1.15}
                         \begin{tabular}{|p{14.5cm}|}
                         \hline}
\newcommand{\deft}      {\tabt{Definition}}
\newcommand{\defst}     {\tabt{Definitions}}
\newcommand{\ext}       {\tabt{Example}}
\newcommand{\tabc}      {\hline
                         \end{tabular}
                         \vspace{0.1cm}
                         \renewcommand{\arraystretch}{1.15}
                         \begin{tabular}{|p{14.5cm}|}
                         \hline}
\newcommand{\tabb}      {\hline
                         \end{tabular}
                         \vspace{0.2cm}
                         \end{flushleft}}
\begin{document}
\thispagestyle{empty}
\begin{flushright}
 October 2009
\end{flushright}
\begin{center}
{\Large\bf
 Uploading User--Defined Functions \\ \vspace{0.2cm}
 onto the \amidas\ Website}        \\
\vspace*{0.7cm}
 {\sc Chung-Lin Shan} \\
\vspace*{0.5cm}
 {\it School of Physics and Astronomy, Seoul National University \\
      Seoul 151-747, Republic of Korea}                          \\
%\vspace*{0.1cm}
%
 {\it E-mail:} {\tt cshan@hep1.snu.ac.kr}
\end{center}
\vspace{1cm}
\begin{abstract}
 The \amidas\ website
 has been established as an online interactive tool
 for running simulations and analyzing data
 in direct Dark Matter detection experiments.
 At the first phase of the website building,
 only some commonly used WIMP velocity distribution functions
 and elastic nuclear form factors
 have been involved in the \amidas\ code.
 In order to let the options for velocity distribution
 as well as for nuclear form factors
 be more flexible,
 we have extended the \amidas\ code to be able to
 include {\em user--uploaded} files
 with their own functions.
 In this article,
 I describe the preparation of files of
 user--defined functions onto the \amidas\ website.
 Some examples will also be given.
\end{abstract}
\clearpage
\section{Introduction}
 In the last few years
 we developed new methods
 for analyzing data, i.e., measured recoil energies,
 from (future) direct Dark Matter detection experiments
 as model--independently as possible
 \cite{DMDDf1v, DMDDmchi,
       DMDDfp2-IDM2008, DMDDidentification-DARK2009}.
 These methods will help us to
 understand the nature of WIMP
 (Weakly Interacting Massive Particle $\chi$)
 Dark Matter,
 to identify them among new particles
 produced hopefully in the near future at colliders,
 as well as to reconstruct
 the (sub)structure of our Galactic halo.
 Following the development of
 these model--independent data analysis procedures,
 we combined the programs for simulations to a compact system:
 \amidas\ (A Model--Independent Data Analysis System).
 For users' convenience and
 under the collaboration with the ILIAS Project \cite{ILIAS},
 an online system has also been established
 at the same time \cite{AMIDAS-web, AMIDAS-SUSY09}.

 For the first version of the \amidas\ code and website,
 the options for target nuclei,
 for the velocity distribution function of halo WIMPs,
 as well as for the elastic nuclear form factors
 for spin--independent (SI) and spin--dependent (SD)
 WIMP--nucleus interactions
 are fixed and only some commonly used forms
 have been involved in the \amidas\ code
 \cite{AMIDAS-SUSY09}.
 Users can not choose different detector materials
 nor use different velocity distribution/form factors
 for their simulations and/or data analyses.
 In order to let the options for the velocity distribution
 as well as for the nuclear form factors
 be more flexible,
 we have extended the \amidas\ code to be able to
 include {\em user--uploaded} files
 with their own functions.
 Note that,
 since the \amidas\ code has been written
 in the C programming language,
 all user--defined functions for uploading
 must also be given using the syntax of C.

 The remainder of this article is organized as follows.
 In Sec.~2
 I will talk about setting users' own target nuclei.
 In Secs.~3 and 4
 the preparation of files defining
 the velocity distribution function
 and
 the nuclear form factors
 will be described,
 respectively.
 I will conclude in Sec.~5.
 Some intrinsically defined constants and functions
 in the \amidas\ code will be given in an appendix.
\section{Target nuclei}
 Users can give as usual the element symbol
 and the atomic mass number $A$ of their chosen target nuclei
 on the website directly.
 For the determination of ratios of
 different WIMP--nucleon couplings/cross sections,
 the total spin of the target nucleus $J$
 and the expectation values of the proton and neutron group spins
 $\expv{S_{\rm (p, n)}}$
 for the chosen targets with spin sensitivities
 are also required.
\section{Velocity distribution of halo WIMPs}
 For defining the one--dimensional WIMP velocity distribution,
 users should in practice give
 the integral over the velocity distribution function:
\beq
   \int_{\vmin = \alpha \sqrt{Q}}^{\vmax} \bfrac{f_1(v, t)}{v} dv
\eeq
 with the name of ``\verb+Intf1v_v_user+''.
 Note here that
 the lower limit of the integral,
 the minimal incoming velocity of incident WIMPs
 that can deposit the energy $Q$ in the detector, $\vmin$,
 must be expressed as a function of the energy $Q$ through
 $\vmin = \alpha \sqrt{Q}$;
 the upper limit of the integral, $\vmax$,
 is often set as the escape velocity $\vesc$ or even $\infty$,
 since the WIMP flux on the Earth
 is usually assumed to be negligible
 at velocities $v \ge \vesc~\gsim~600$ km/s.
 Meanwhile,
 since 
\beq
        \alpha
 \equiv \sfrac{\mN}{2 \mrN^2}
%\~,
\label{eqn:alpha}
\eeq
 with the reduced mass
\beq
        \mrN
 \equiv \frac{\mchi \mN}{\mchi + \mN}
%\~,
\label{eqn:mrN}
\eeq
 is a function of the WIMP mass $\mchi$
 and the mass of target nucleus $\mN$,
 which has been defined as a function of the atomic mass number $A$,
 the expression of the integral over the WIMP velocity distribution
 should also be a function of $\mchi$ and $A$.
 Finally,
 the velocity distribution function and
 thus the expression of the integral
 should generally be a function of time $t$.

 As examples,
 the integral over the simple
 Maxwellian velocity distribution function
 \cite{SUSYDM96, DMDDf1v} can be defined as
\ext
\verb# double Intf1v_v_user(double mchi, int A, double QQ, double tt) # \\
\verb# {                                                              # \\
\verb#   return                                                       # \\
\verb#     (2.0 / sqrt(M_PI) / (v_0 * v_U)) *                         # \\
\verb#     exp(-alpha(mchi, A) * alpha(mchi, A) * QQ /                # \\
\verb#          (  (v_0 * v_U) * (v_0 * v_U)  )       );              # \\
\verb# }                                                              # \\
\tabb
 and the integral over the shifted Maxwellian velocity distribution
 \cite{SUSYDM96, DMDDf1v} can be given by
\ext
\verb# double Intf1v_v_user(double mchi, int A, double QQ, double tt)               # \\
\verb# {                                                                            # \\
\verb#   return                                                                     # \\
\verb#     (1.0 / 2.0 / (v_e(tt) * v_U)) *                                          # \\
\verb#     (  erf( (alpha(mchi, A) * sqrt(QQ) + (v_e(tt) * v_U)) / (v_0 * v_U) )    # \\
\verb#      - erf( (alpha(mchi, A) * sqrt(QQ) - (v_e(tt) * v_U)) / (v_0 * v_U) ) ); # \\
\verb# }                                                                            # \\
\tabb
 Here \verb+v_e(tt)+,
 standing for the {\em time--dependent} Earth's velocity
 in the Galactic frame $\ve(t)$,
 has been defined in the \amidas\ code by
\deft
\verb# double v_e(double tt)                                     # \\
\verb# {                                                         # \\
\verb#   return v_0 * ( 1.05 + 0.07 * cos(omega * (tt - t_p)) ); # \\
\verb# }                                                         # \\
\tabb
 Note that,
 firstly,
 \verb+v_U+,
 the function \verb+alpha(mchi, A)+,
 where \verb+mchi+ and \verb+A+ stand for
 the WIMP mass $\mchi$ and
 the atomic mass number of the target nucleus, $A$,
 and the constant \verb+omega+
 are defined intrinsically in the \amidas\ code.
 Secondly,
 \verb+v_0+ and \verb+t_p+, standing for
 the Sun's orbital velocity in the Galactic frame, $v_0$,
 and the date on which
 the velocity of the Earth relative to the WIMP halo is maximal, $t_{\rm p}$,
 are two input parameters
 which users can set later on the website
 \cite{AMIDAS-SUSY09}.
 Thirdly,
 \verb+QQ+ and \verb+tt+ stand for
 the energy variable $Q$ and time variable $t$.
 Finally,
 for a {\em time--independent} velocity distribution
 or a numerical expression with
 a fixed experimental running time $t = t_{\rm expt}$,
 ``\verb+double tt+'' should still be declared
 as one of the {\em four} variables of the function \verb+Intf1v_v_user+.  

 On the other hand,
 for comparing results of the reconstructed velocity distribution
 with the (input) theoretical one in the output plots, 
 users need an {\em extra file}
 for giving the velocity distribution function {\em itself}
 (not the integral over it now)
 for drawing the (input) theoretical
 velocity distribution function $f_1(v)$.
 Since the {\tt gnuplot} software \cite{gnuplot}
% in the UNIX system
 has been used for drawing output plots,
 the velocity distribution function
 given in this file
 must be written using the syntax of {\tt gnuplot}
 with the name of ``\verb+f1v_user(x)+''.
 As examples,
 the simple Maxwellian velocity distribution
 can be given as
\ext
\verb#   M_PI = 3.1416                         # \\
\verb#                                         # \\
\verb#   v_0 = 220.0                           # \\
\verb#                                         # \\
\verb#      f1v_user(x)                      \ # \\
\verb#   =  (4.0 / sqrt(M_PI)) *             \ # \\
\verb#      ((x * x) / (v_0 * v_0 * v_0)) *  \ # \\
\verb#      exp(-(x * x) / (v_0 * v_0))      \ # \\
\tabb
 and for the shifted Maxwellian velocity distribution:
\ext
\verb#   M_PI = 3.1416                                             # \\
%\verb#                                                             # \\
\verb#   omega  = 2.0 * M_PI / 365.0                               # \\
\verb#                                                             # \\
\verb#   t_p    = 152.5                                            # \\
\verb#   t_expt = 243.75                                           # \\
\verb#                                                             # \\
\verb#   v_0 = 220.0                                               # \\
\verb#   v_e = v_0 * ( 1.05 + 0.07 * cos(omega * (t_expt - t_p)) ) # \\
\verb#                                                             # \\
\verb#     f1v_user(x)                                           \ # \\
\verb#  =  1.0 / sqrt(M_PI) * (x / v_e / v_0) *                  \ # \\
\verb#     (  exp(-(x - v_e) * (x - v_e) / (v_0 * v_0))          \ # \\
\verb#      - exp(-(x + v_e) * (x + v_e) / (v_0 * v_0))  )       \ # \\
\tabb
 Note here that
 the ``\verb+\+'' (backslash) must be used
 (also at the end of the last line!)
 in order to insert the definition given in this file
 into the other intrinsic commands correctly.

 Two sample files,
 one is for the \amidas\ code and
 the other one is for the {\tt gnuplot} software,
 can be downloaded from the \amidas\ website.
\section{Nuclear form factors}
 For defining users' own nuclear form factors,
 {\em not only} the definition of the {\em squared} form factor $\FQ$
 {\em but also} its derivative with respect to the energy $Q$:
\beq
   \Dd{\FQ}{Q}
 = 2 F(Q) \bDd{F(Q)}{Q}
%\~,
\label{eqn:dFQdQ}
\eeq
 (not $F(Q)$ itself nor $dF(Q) / dQ$)
 must be given in {\em one} file together
 with the names of ``\verb+FQ_SI_user+ (\verb+FQ_SD_user+)''
 and ``\verb+dFQdQ_SI_user+ (\verb+dFQdQ_SD_user+)''.
 The expression of the squared form factor
 can be either a general analytic form for different nuclei
 or a specified form for the chosen nucleus.
 As examples,
 for the SI WIMP--nucleus cross section,
 the exponential form factor $F_{\rm ex}^2(Q)$
 \cite{Ahlen87, Freese88, SUSYDM96}
 can be given as
\ext
\verb# double FQ_SI_user(int A, double QQ) # \\
\verb# {                                   # \\
\verb#   return exp(-QQ / Q_0(A));         # \\
\verb# }                                   # \\
%\tabb
%
%\tabc
\verb#                                               # \\
%
%\ext
\verb# double dFQdQ_SI_user(int A, double QQ)        # \\
\verb# {                                             # \\
\verb#   return -(1.0 / Q_0(A)) * FQ_SI_user(A, QQ); # \\
\verb# }                                             # \\
\tabb
 and the Woods--Saxon form factor $F_{\rm WS}^2(Q)$
 \cite{Engel91, SUSYDM96}:
\ext
\verb# double FQ_SI_user(int A, double QQ)                                        # \\
\verb# {                                                                          # \\
\verb#   if (QQ == 0.0)                                                           # \\
\verb#     return 1.0;                                                            # \\
\verb#                                                                            # \\
\verb#   else                                                                     # \\
\verb#   {                                                                        # \\
\verb#     return                                                                 # \\
\verb#       ( 3.0 * sphBesselj(1, qq(A, QQ) * R_1(A)) / (qq(A, QQ) * R_1(A)) ) * # \\
\verb#       ( 3.0 * sphBesselj(1, qq(A, QQ) * R_1(A)) / (qq(A, QQ) * R_1(A)) ) * # \\
\verb#       exp(-(qq(A, QQ) * ss) * (qq(A, QQ) * ss));                           # \\
\verb#   }                                                                        # \\
\verb# }                                                                          # \\
%\tabb
%
%\tabc
\verb#                                                         # \\
%
%\ext
\verb# double dFQdQ_SI_user(int A, double QQ)                  # \\
\verb# {                                                       # \\
\verb#   if (QQ == 0.0)                                        # \\
\verb#     return -0.4 * (R_A(A) * R_A(A)) * m_N(A);           # \\
\verb#                                                         # \\
\verb#   else                                                  # \\
\verb#   {                                                     # \\
\verb#     return                                              # \\
\verb#       (  dsphBesselj(1, qq(A, QQ) * R_1(A)) * R_1(A) /  # \\
\verb#           sphBesselj(1, qq(A, QQ) * R_1(A))             # \\
\verb#        - 1.0 / qq(A, QQ)                                # \\
\verb#        - (qq(A, QQ) * ss * ss) ) *                      # \\
\verb#       FQ_SI_user(A, QQ) * ((2.0 * m_N(A)) / qq(A, QQ)); # \\
\verb#   }                                                     # \\
\verb# }                                                       # \\
\tabb
 Here the spherical Bessel functions
\beq
   j_n(x)
 = \sfrac{2}{\pi x} \~ J_{n + 1/2}(x)
\~,
\eeq
 their derivatives
\beq
   j_n'(x)
 = \sfrac{2}{\pi x}
   \bbrac{-\frac{1}{2 x} \~ J_{n + 1/2}(x) + J_{n + 1/2}'(x)}
\~,
\eeq
 as well as the (derivatives of the) half--integer
 Bessel functions $J_{n + 1/2}(x)$,
 for $n = 0,~\pm 1,~\pm 2,~\cdots$,
 are given in an intrinsic package of the \amidas\ code.
 The functions
 \verb+Q_0(A)+, \verb+qq(A, QQ)+, \verb+R_1(A)+, \verb+R_A(A)+, \verb+m_N(A)+,
 and the constant \verb+ss+
 are also defined intrinsically in the \amidas\ code
 (see the appendix).

 As a more complicated example,
 the thin--shell form factor $F_{\rm TS}^2(Q)$
 used sometimes for the SD WIMP--nucleus cross section
 \cite{Lewin96, Klapdor05}
 can be defined as
\ext
\verb# double FQ_SD_user(int A, double QQ)                                          # \\
\verb# {                                                                            # \\
\verb#   if (QQ == 0.0)                                                             # \\
\verb#     return 1.0;                                                              # \\
\verb#                                                                              # \\
\verb#   else                                                                       # \\
\verb#   if (QQ <= QQ_SD_min(A) ||                                                  # \\
\verb#       QQ >= QQ_SD_max(A)   )                                                 # \\
\verb#   {                                                                          # \\
\verb#     return                                                                   # \\
\verb#       sphBesselj(0, qq(A, QQ) * R_1(A)) * sphBesselj(0, qq(A, QQ) * R_1(A)); # \\
\verb#   }                                                                          # \\
\verb#                                                                              # \\
\verb#   else                                                                       # \\
\verb#     return FQ_TS_const;                                                      # \\
\verb# }                                                                            # \\
%\tabb
%
%\tabc
\verb#                                                                          # \\
%
%\ext
\verb# double dFQdQ_SD_user(int A, double QQ)                                   # \\
\verb# {                                                                        # \\
\verb#   if (QQ == 0.0)                                                         # \\
\verb#     return -(1.0 / 3.0) * (R_1(A) * R_1(A)) * (2.0 * m_N(A));            # \\
\verb#                                                                          # \\
\verb#   else                                                                   # \\
\verb#   if (QQ <= QQ_SD_min(A) ||                                              # \\
\verb#       QQ >= QQ_SD_max(A)   )                                             # \\
\verb#   {                                                                      # \\
\verb#     return                                                               # \\
\verb#       dsphBesselj(0, qq(A, QQ) * R_1(A)) * R_1(A) *                      # \\
\verb#        sphBesselj(0, qq(A, QQ) * R_1(A)) * ((2.0 * m_N(A)) / qq(A, QQ)); # \\
\verb#   }                                                                      # \\
\verb#                                                                          # \\
\verb#   else                                                                   # \\
\verb#     return 0.0;                                                          # \\
\verb# }                                                                        # \\
\tabb
 The functions
 \verb+QQ_SD_min(A)+, \verb+QQ_SD_max(A)+,
 and the constant \verb+FQ_TS_const+
 are defined in the \amidas\ code (see the appendix).
\section{Summary}
 In this article,
 I described the preparation of files
 giving user--defined WIMP velocity distribution function
 and/or elastic nuclear form factor(s)
 which can be uploaded onto the \amidas\ website
 for more flexible simulations or data analyses.
 This improvement allows theorists
 to simulate with their own/favorite models and
 compare them with (future) experimental results,
 as well as gives experimentalists flexible choices
 for more suitable form factor(s) for their own detector materials.

 In summary,
 up to now all basic functions of
 the \amidas\ code and website
 have been well established.
 Hopefully this new tool can help our colleagues
 to detect/discover WIMP Dark Matter,
 to understand the nature of Dark Matter particles
 and the (sub)structure of the Galactic halo in the future.
\subsubsection*{Acknowledgments}
 The author would like to thank
 James Y.~Y.~Liu
 for discussing the basic idea
 of combining user--defined functions
 with the source code.
 The author would be grateful to the ILIAS Project and
 the Physikalisches Institut der Universit\"at T\"ubingen
 for kindly providing the opportunity of the collaboration
 and the technical support of the \amidas\ website.
 The author would also like to thank
 the friendly hospitality of the Institute of Physics,
 National Chiao Tung University
 where part of this work was completed.
 This work was partially supported
 by the BK21 Frontier Physics Research Division under project
 no.~BA06A1102 of Korea Research Foundation.
\appendix
\setcounter{equation}{0}
\setcounter{figure}{0}
\renewcommand{\theequation}{A\arabic{equation}}
\renewcommand{\thefigure}{A\arabic{figure}}
%
%
% Appendix A
%
\section{Relevant constants and functions defined in the \amidas\ code}
 Here I list some relevant constants and functions
 defined intrinsically in the \amidas\ code,
 which could be needed by users
 for defining their own functions.
\subsection{Defined constants}
 The constants defined in the \amidas\ code
 are in {\em natural units} as following:
%
% c = 2.997 924 58 * 10^5 km / s
% hbar * c = 0.197 326 968(17) GeV fm
%
\defst
\verb# double c = 1.0;                    # \\
\verb# double v_U = c / (2.998 * 1e5);    # \\
\verb# double m_U = 1e6 / (c * c);        # \\
\verb# double fm_U = c / (0.1973 * 1e6);  # \\
\verb# double m_p = 0.938 * m_U;          # \\
\verb# double omega = 2.0 * M_PI / 365.0; # \\
\tabb
\subsection{Defined functions}
 The functions used for defining
 the commonly used WIMP velocity distribution functions
 and nuclear form factors in the \amidas\ code
 are given here.
\subsubsection{The reduced mass \boldmath$\mrN(\mchi, A)$
               and $\alpha(\mchi, A)$}
 All masses in the \amidas\ code are in the unit of GeV$/c^2$.
 Firstly,
 the mass of the target nucleus
 is defined as a function of the atomic mass number $A$ by
\deft
\verb# double m_N(int A)        # \\
\verb# {                        # \\
\verb#   return m_p * A * 0.99; # \\
\verb# }                        # \\
\tabb
 Here the mass difference between a proton and a neutron
 has been neglected.
 And the reduced mass between the WIMP mass
 and ``something'' is given by
\deft
\verb# double mchi_r(double mchi, double mx) # \\
\verb# {                                     # \\
\verb#   return mchi * mx / (mchi + mx);     # \\
\verb# }                                     # \\
\tabb
 Therefore,
 the reduced mass of the WIMP mass
 with the mass of the target nucleus,
 $\mrN$ in Eq.(\ref{eqn:mrN}),
 and with the proton mass
 are defined as
\defst
\verb# double mchi_rN(double mchi, int A)   # \\
\verb# {                                    # \\
\verb#   return mchi_r(mchi * m_U, m_N(A)); # \\
\verb# }                                    # \\
%\tabb
%
% and
\verb#                                   # \\
%
%\deft
\verb# double mchi_rp(double mchi)       # \\
\verb# {                                 # \\
\verb#   return mchi_r(mchi * m_U, m_p); # \\
\verb# }                                 # \\
\tabb
 Finally,
 $\alpha$ defined in Eq.(\ref{eqn:alpha})
 has been given as a function of $\mchi$ and $A$:
\deft
\verb# double alpha(double mchi, int A)                # \\
\verb# {                                               # \\
\verb#   return sqrt(m_N(A) / 2.0) / mchi_rN(mchi, A); # \\
\verb# }                                               # \\
\tabb
\subsubsection{For nuclear form factors}
 For the exponential form factor $F_{\rm ex}(Q)$,
 the nuclear radius $R_0$
 and the nuclear coherence energy $Q_0$
 have been defined
 as functions of the atomic mass number $A$:
\defst
\verb# double R_0(int A)                                  # \\
\verb# {                                                  # \\
\verb#   return (0.3 + 0.91 * cbrt(m_N(A) / m_U)) * fm_U; # \\
\verb# }                                                  # \\
%\tabb
%
% and
%\verb#                                            # \\
\tabc
%
%\deft
\verb# double Q_0(int A)                          # \\
\verb# {                                          # \\
\verb#   return 1.5 / (m_N(A) * R_0(A) * R_0(A)); # \\
\verb# }                                          # \\
\tabb
 For the Woods--Saxon form factor $F_{\rm WS}(Q)$,
 the nuclear skin thickness $s$,
 the effective nuclear radius $R_A$,
 and the radius $R_1$
 are defined as
\defst
\verb# double ss = fm_U; # \\
%\tabb
%
\verb#                                # \\
%
%\deft
\verb# double R_A(int A)              # \\
\verb# {                              # \\
\verb#   return 1.2 * cbrt(A) * fm_U; # \\
\verb# }                              # \\
%\tabb
%
\verb#                                # \\
%
%\deft
\verb# double R_1(int A)                               # \\
\verb# {                                               # \\
\verb#   return sqrt(R_A(A) * R_A(A) - 5.0 * ss * ss); # \\
\verb# }                                               # \\
\tabb
 Meanwhile,
 the transferred 3--momentum $q = \sqrt{2 \mN Q}$ has been give
 as a function of the atomic mass number $A$ and recoil energy $Q$:
\deft
\verb# double qq(int A, double QQ)       # \\
\verb# {                                 # \\
\verb#   return sqrt(2.0 * m_N(A) * QQ); # \\
\verb# }                                 # \\
\tabb
 For the thin--shell form factor $F_{\rm TS}(Q)$,
 the dimensionless constants
 giving the lower and upper energy bounds,
 between which the form factor is a constant,
 and this constant ($\simeq 0.047$)
 are given as
\defst
\verb# double qqR_1_min = 2.55;                                           # \\
\verb# double qqR_1_max = 4.50;                                           # \\
\verb#                                                                    # \\
\verb# double FQ_TS_const;                                                # \\
\verb# FQ_TS_const = sphBesselj(0, qqR_1_min) * sphBesselj(0, qqR_1_min); # \\
\tabb
 Then the lower and upper energy bounds
 can be estimated by
\defst
\verb# double QQ_SD_min(int A)                                              # \\
\verb# {                                                                    # \\
\verb#   return (qqR_1_min * qqR_1_min) / (2.0 * m_N(A) * R_1(A) * R_1(A)); # \\
\verb# }                                                                    # \\
%\tabb
%
\verb#                                                                      # \\
%
%\deft
\verb# double QQ_SD_max(int A)                                              # \\
\verb# {                                                                    # \\
\verb#   return (qqR_1_max * qqR_1_max) / (2.0 * m_N(A) * R_1(A) * R_1(A)); # \\
\verb# }                                                                    # \\
\tabb
\subsection{Declared variables}
 Here I list the {\em short used names}
 for variables and constants in the \amidas\ code.
 Note that
 these names should be {\em avoided} to use
 in the user--uploaded code(s),
 otherwise the \amidas\ program might not work correctly.
\defst
\verb# AX, AX_str, Ax, calN, fp_U, G_F, hn, InX, JJ, k_1, ka, kn, lambda, ln,       # \\
\verb# mchi_assumed, mchi_tmp, mm, m_N_TX, nexpt, nmchi, NmX, nn, nn_max, nn_win,   # \\
\verb# nplot, nranap, nTX, pb_U, pm, Qmax, Qmid, Qmin, Qwin, Qvin, Qtp, rho, rho_0, # \\
\verb# rho_U, RJ, RJX, rlh, Rn, Rn_sh, RnX, Rsigma, R_TX, run, runi, runj, runk,    # \\
\verb# r_win, sh, Sn, Snp, sol, Sp, Spn, std_rho_0, std_t_p, std_v_0, s_U, t_end,   # \\
\verb# t_expt, tmax, tmid, tmin, t_start, TX, v_e_annual, v_esc    # \\
\tabb

\begin{thebibliography}{99}
%
\bibitem{DMDDf1v}
 {M.~Drees and C.~L.~Shan,
%  {\it ``Reconstructing the Velocity Distribution of Weakly Interacting Massive Particles
%         from Direct Dark Matter Detection Data''},
  {\it J.~Cosmol.~Astropart.~Phys.} {\bf 0706}, 011 (2007).}
%
\bibitem{DMDDmchi}
 {M.~Drees and C.~L.~Shan,
%  {\it ``Model--Independent Determination of the WIMP Mass
%         from Direct Dark Matter Detection Data''},
  {\it J.~Cosmol.~Astropart.~Phys.} {\bf 0806}, 012 (2008).}
%
\bibitem{DMDDfp2-IDM2008}
 {M.~Drees and C.~L.~Shan,
%  {\it ``Constraining the Spin--Independent WIMP--Nucleon Coupling
%         from Direct Dark Matter Detection Data''},
  {\it proceedings of IDM 2008},
%       the 7th International Workshop on the Identification of Dark Matter (IDM 2008)},
  {\tt arXiv:0809.2441 [hep-ph]} (2008).}
%
\bibitem{DMDDidentification-DARK2009}
 {M.~Drees and C.~L.~Shan,
%  {\it ``How Precisely Could We Identify WIMPs Model--Independently
%         with Direct Dark Matter Detection Experiments''},
  {\it proceedings of DARK 2009},
%       the Seventh International Heidelberg Conference
%       on Dark Matter in Astro and Particle Physics (DARK 2009)},
  {\tt arXiv:0903.3300 [hep-ph]} (2009).}
%
\bibitem{ILIAS}
 {% The ILIAS Project,
  {\tt http://www-ilias.cea.fr/}.}
%
\bibitem{AMIDAS-web}
 {%The \amidas\ (A Model--Independent Data Analysis System)
  %code and website for direct Dark Matter detection experiments, \\
  {\tt http://pisrv0.pit.physik.uni-tuebingen.de/darkmatter/amidas/}.}
%
\bibitem{AMIDAS-SUSY09}
 {C.~L.~Shan,
%  {\it ``The {\cmitttxii AMIDAS} Website: An Online Tool for
%         Direct Dark Matter Detection Experiments''},
  {\it proceedings of SUSY09},
%       the 17th International Conference on
%       Supersymmetry and the Unification of Fundamental Interactions (SUSY09)},
%  {\it AIP Conf.~Proc.} {\bf 1xxx}, xxx (2010),
  {\tt arXiv:0909.1459 [astro-ph.IM]} (2009).}
%
\bibitem{SUSYDM96}
 {G.~Jungman, M.~Kamionkowski, and K.~Griest,
%  {\it ``Supersymmetric Dark Matter''},
  {\it Phys.~Rep.} {\bf 267}, 195 (1996).}
%
\bibitem{gnuplot}
 {% The gnuplot homepage,
  {\tt http://www.gnuplot.info/}.}
%
% Exponential form factor
\bibitem{Ahlen87}
 {S.~P.~Ahlen {\it et al.},
  {\it Phys.~Lett.} {\bf B 195}, 603 (1987).}
%
\bibitem{Freese88}
 {K.~Freese, J.~Frieman, and A.~Gould,
  {\it Phys.~Rev.} {\bf D 37}, 3388 (1988).}
%
% Woods-Saxon form factor
\bibitem{Engel91}
 {J.~Engel,
%  {\it ``Nuclear Form--Factors for the Scattering of
%         Weakly Interacting Massive Particles''},
  {\it Phys.~Lett.} {\bf B 264}, 114 (1991).}
%
\bibitem{Lewin96}
 {J.~D.~Lewin and P.~F.~Smith,
%  {\it ``Review of Mathematics, Numerical Factors, and Corrections
%         for Dark Matter Experiments Based on Elastic Nuclear Recoil''},
  {\it Astropart.~Phys.} {\bf 6}, 87 (1996).}
%
% Thin-shell model
\bibitem{Klapdor05}
 {H.~V.~Klapdor-Kleingrothaus, I.~V.~Krivosheina, and C.~Tomei,
  {\it Phys.~Lett.} {\bf B 609}, 226 (2005).}
%
\end{thebibliography}
\end{document}